\documentclass[useAMS,usenatbib]{mn2e}
\usepackage{amsfonts}
\usepackage{graphicx}

\newcommand{\ha}{\hbox{H$\alpha$}}
\newcommand{\etal}{{et al.}}

\title[The CIDA-UCM-Yale Shallow Survey for Emission Line Galaxies]
{The CIDA-UCM-Yale Shallow Survey for Emission Line Galaxies}
\author[A. Bongiovanni, G. Bruzual, G. Magris, J. Gallego, C.E. Garc{\'\i}a-Dab\'o, P. Coppi \\
~and C. Sabbey]
{A. Bongiovanni$^{1}$\thanks{E-mail: abongiov@cida.ve (ABP); bruzual@cida.ve (GBA)}, 
G. Bruzual$^{1}$, G. Magris$^{1}$, J. Gallego$^{2}$, C.E. Garc{\'\i}a-Dab\'o$^{2}$, 
\newauthor P. Coppi$^{3}$, C. Sabbey$^{4}$\\
\\
$^{1}$Centro de Investigaciones de Astronom{\'\i}a, CIDA, AP 264, M\'erida 5101-A,
Venezuela\\
$^{2}$Universidad Complutense de Madrid, Departamento de Astrof{\'\i}sica y Ciencias de la Atm\'osfera,
Facultad de Ciencias F{\'\i}sicas,\\
Madrid, Espa\~na\\
$^{3}$Yale University, Astronomy Department, P.O. Box 208101, New Haven, CT 06520, USA\\
$^{4}$Bogle Investment Management, Wellesley, MA 02481, USA\\}

\begin{document}

\date{Accepted ... Received ...; in original form ...}

\pagerange{\pageref{firstpage}--\pageref{lastpage}} \pubyear{2004}

\maketitle

\label{firstpage}

\begin{abstract}
We present the CIDA-UCM-Yale (Centro de Investigaciones de Astronom\'{\i}a, Universidad
Complutense de Madrid and Yale University) survey for
$\ha+[{\rm NII}]6549,6584$ emission-line galaxies using objective-prism spectra. The
most important properties of a catalogue with 427 entries and significant subsets are 
analysed. The complete sample contains 183 statistically confirmed ELGs in a sky area
of 151 deg$^2$ and redshift up to 0.14. We determine the parameters of the $\ha$ 
luminosity function using the $\ha+[{\rm NII}]$ flux directly measured on the
ELGs spectra in this sample and the star formation rate density derived
is in agreement with the values reported in the literature. Finally, we study the 
clustering properties of local star-forming galaxies relative to quiescent ones from 
different perspectives. We find that emission-line galaxies avoid dense regions of
quiescent galaxies and we propose a power-law expression to parametrise the relation
between star formation rate density and environment volume density of emission-line
galaxies.
\end{abstract}


\section{Introduction}

The star formation rate density ($\rho_{\rm SFR}$) in the Local Universe and
the spatial distribution of emission-line galaxies (ELGs) as a function of
environment are fundamental pieces in the field of formation and evolution
of galaxies. There are few techniques as good as slitless spectroscopy for 
bulk-searching of this object class. From the mid 1960s to the near past the
photographic plates were used as unique detectors in the search on ELGs
({\rm e.g.}  \citealt{1977ApJS...35..197M}; \citealt{1983ApJ...272...68W}; 
\citealt{1983ApJS...51..171P}; \citealt{1986Afz....25..345M};
\citealt{1996A&AS..116...43P}; \citealt{1998A&AS..133..171S} and references therein;
\citealt{1999A&AS..135..511U}). The UCM (Universidad Complutense de Madrid) survey 
\citep{1994ApJS...95..387Z}
is one of the more recent initiatives devoted to the photographic searching of ELGs 
and their findings motivated this work. Substitution of photographic plates
by large-format CCDs, including CCD mosaics, is really a challenge for many
observatories with competitive aperture Schmidt telescopes, some of them
equipped with objective-prisms. Consequently, the CCD-based objective-prism surveys
could be taken as a promising technique for the search of extragalactic emission-line 
systems. Examples of two successful objective-prism digital surveys are presented
by \citep{2001ApJ...548..585S} and \citep{2004AJ....127.1943G}, the latter best 
known as the KPNO international spectroscopic survey, KISS. 

The CIDA-UCM-Yale (Centro de Investigaciones de Astronom\'{\i}a, Universidad
Complutense de Madrid and Yale University)
survey\footnotemark[1],
\footnotetext[1]{The data presented here were partially extracted from the
QUEST1 Collaboration raw data repository. QUEST1 is short for the Quasar
Equatorial Survey Team, and it was the first collaboration between groups from
Yale University, Indiana University, Centro de Investigaciones de
Astronom\'{\i}a (CIDA), and Universidad de Los Andes (M\'erida, Venezuela).}
hereafter CUYS, is the first stage of an observational effort with the main goal
of identifying and studying emission-line galaxies (ELGs) based on digital 
low-resolution spectra in contrast with spectroscopic surveys which involve 
follow-up
observations or multifiber instruments. Once the second part of this 
equatorial objective-prism survey is completed, we expect
to have catalogued more than 10,000 ELGs with blue and red low-resolution
spectra (4,000-9,200 \AA) over $\sim600\ $deg$^2$.
The CUYS is being carried out 
in the 1m Schmidt telescope located at the Venezuelan National
Astronomical Observatory, Llano del Hato, M\'erida, Venezuela \footnotemark[2]. 
This telescope
\footnotetext[2]{Operated by CIDA and funded by the Ministerio de 
Ciencia y Tecnolog\'{\i}a, Venezuela.}
is equipped with a $3.4\degr$ objective-prism (reciprocal dispersion 
$\sim$25\ \AA\ arcsec$^{-1}$ in $\ha$ at rest)
and a $4\times4$ CCD mosaic covering
$2.3\degr\times3.5\degr$ on the sky. The CCDs are $2048\times2048$
LORAL devises with 15$\mu$m pixels. The camera and its operation details
are fully described in \cite{2002PASP....114..780}. 

Our strategy to select ELGs is
the detection of $\ha+[{\rm NII}]6549,6584$\ \AA\ emission on reduced 1-dimensional
spectra of extended sources listed in the APM catalogue \citep{1990MNRAS....243..692}.
Because objective-prism dispersion decreases in the red portion of the spectra, absorption 
bands present in late-type star spectra could generate false emission
line detections. To avoid these contaminants, among others, we chose to 
ignore the spectra of point-like sources. Taking
into account the galactic latitude range of the CUYS 
observations, data reduction allowed us to obtain a statistically complete
catalogue of active star-forming galaxies and to study the spatial 
distribution of these objects on medium and large scales (above $\sim 3$ Mpc).
Several approaches have been used to study this topic during the last 15 years.
It is well established that ELGs can be found near structures defined by
luminous normal galaxies \citep{1989ApJ...347..152S} or populating low galaxy density
regions in the nearby universe \citep{1997A&A...325..881P}.
Consequently, there is a relation between mean star formation rate (SFR) 
and environment in the sense that
$\rho_{\rm SFR}$ decreases as the galaxy number density increases
({\rm e.g.} \citealt{1978MNRAS.183..633G},
\citealt{1985ApJ...288..481D}, \citealt{2002MNRAS.334..673L}). Recently,
\cite{2003ApJ...584..210G} have established an inverse relation between environmental
surface density
and SFR measured using ELGs from the SDSS. A plausible explanation is that
the high mean SFR in low density environments of the local universe is only
a contrast effect resulting from the mild SFR in high density regions
in which galaxies are stripped of their gas with more probability.

This paper is organized as follows.
An analysis of the catalogue properties is the aim of section 2.
In section 3 we use the resulting CUYS ELG catalogue to find the parameters of the 
\ha\ Luminosity Function
(LF) and estimate the $\rho_{\rm SFR}$
in the Local Universe using fluxes and
equivalent widths ($EW$) measured directly from objective-prism spectra.
In section 4 
we analyse the spatial distribution of ELGs in our survey
in terms of the 2-point correlation function
and other statistical estimators related to the spatial distribution at large
scales. We conclude in section 5. Throughout this paper, 
we adopt the standard notation $h=H_0/100\ {\rm km}\ {\rm s}^{-1}\ {\rm Mpc}^{-1}$
and the cosmology $\Omega_m=0.3$
and $\Omega_\Lambda=0.7$, nevertheless the latter assumption 
has not much effect on the redshift range of our survey.

\begin{table*}
\centering
\begin{minipage}{170mm}
\caption{\label{t:catalogue}List of the first 10 entries in the CUYS catalogue. 
Right ascension, declination and R magnitude were extracted from the USNO-A V2.0 catalogue. The
redshift, equivalent width [\AA] and \ha+[NII] flux [erg s$^{-1}$ cm$^{-2}$], were
measured directly on the CUYS objective-prism spectra.}
\label{catalogue}
\begin{tabular}{ccccccc} 
\hline
Object Name&R.A.(J2000.0)&Dec.(J2000.0)&R&Redshift&EW(\ha+[NII])&log flux(\ha+[NII])\\ \hline 
CUYS 091452-003359&09:14:52.40& -00:33:58.79& 15.18& 0.060& 25.3&-14.45 \\
CUYS 091955-005953&09:19:55.32& -00:59:52.58& 14.68& 0.097& 32.1&-14.06 \\
CUYS 092244-001853&09:22:44.76& -00:18:52.20& 15.55& 0.057& 17.2&-15.35 \\
CUYS 092439-010238&09:24:39.57& -01:02:37.90& 18.75& 0.302&105.0&-14.04 \\
CUYS 092440-000049&09:24:40.18& -00:00:48.13& 14.35& 0.064& 66.2&-13.90 \\
CUYS 092551-020918&09:25:51.51& -02:09:18.90& 17.66& 0.268& 72.9&-14.01 \\
CUYS 092552-012438&09:25:52.02& -01:24:38.74& 14.91& 0.076& 51.4&-13.96 \\
CUYS 093235-010234&09:32:35.74& -01:02:33.72& 16.44& 0.226& 31.5&-13.94 \\
CUYS 093343-013256&09:33:43.95& -01:32:56.62& 16.13& 0.081& 83.7&-13.79 \\
CUYS 093448-003108&09:34:48.33& -00:31:07.82& 15.92& 0.102& 56.3&-13.90 \\
\hline
\end{tabular}
\end{minipage}
\end{table*}

\section[]{Survey Description}

\subsection{Observations}
Observations were performed in the second half of 1999 and the beginning of
2000 with the instruments
described above. The
use of a red cut-off filter (6300-9200\ \AA) allowed us to minimize 
the overlapping of spectra,
especially in crowded zones, profiting at the same time from the CCDs best response
in this wavelength range. This
instrumental combination limits the $\ha+[NII]$ detection 
in redshift to $z\leq0.4$.  
Data were collected using the driftscan technique, which is also referred to as time-delay
integration mode \citep{1986spie...627..60}. From 23 scans ($\sim0.21$ Tbytes)
we selected the best 18 (which covered repeatedly almost the same regions). We 
detect objects with continuum magnitudes down to $R\sim20$.
These scans
cover the equatorial sky between coordinates $08h\leq\alpha\leq18h$ and 
$-2.4^{\mathrm{o}}\leq\delta\leq+0.1^{\mathrm{o}}$. From the surveyed region, we selected
the best sampled zones, which represent about $250\ $deg$^2$. The different
atmospheric conditions prevailing during the observing
nights and the artificial broadening of the spectra introduced by the driftscan mode 
(produced by residual effects due to the motion of the spectral sagitta and
spread in the rate of motion of spectra across the finite width of a
single CCD, \citealt{2002PASP....114..780}),
result in an average full width at half maximum ({\it fwhm}) of the spectra 
of $\sim3\ $arcsec in the spatial direction.

\subsection{Data Reduction}

To process the observations we used the objective-prism data analysis
package from \cite{1999adass...8..207}, in a version modified by the authors. We omit a detailed
description of this package and only mention the more important steps of 
the analysis. (a) Extraction of the astrometric database from the USNO-A V2.0 
catalogue \citep{1998yCat.1252....0} for object identification; (b) Generation of
calibration files (bias level, flat field vector and bad column list) for each
CCD; (c) Determination of linear coordinate transformation between
the astrometric database and the driftscan strip world coordinates. We used the centroid
of the telluric $A$ band (O$_2$, 7580-7750\ \AA) as a reference to guide the positioning of the spectra in
one of the image axis; (d) Extraction of wavelength calibrated 1-dimensional spectra for 
each object identified
in the USNO-A V2.0 catalogue; (e) Coaddition of 1-dimensional
spectra (four for each driftscan, {\rm i.e.} 560 s of integration time) in a database 
created for this purpose, which also provides information to discard cosmic rays;
(f) Extraction of raw ELG candidates applying an optimal emission line detection
algorithm;
(g) Flux calibration and analysis of ELG candidate spectra
that matched with the previous selection of sources identified as extended in the $b_J-$ 
and $r-$band digitized images of the APM catalogue to discard possible contaminants
in the light of the CUYS goals, {\rm i.e.} late-type stars and QSOs. 
(h) Finally, study of the 1-D spectra of selected candidates, using 
the Image Reduction and Analysis Facility (IRAF\footnotemark[3]) software.
\footnotetext[3]{IRAF is distributed by the National Astronomical Observatory, which
is operated by the Association of Universities for Research in Astronomy, Inc. 
(AURA), under cooperative agreement with the National Science Foundation (USA).}
About this data reduction scheme, the explanations that follow could be useful
to the reader. 

Because the objective-prism 
dispersion decrease in the red portion of the spectrum, the
absorption bands present in late-type stars spectra could generate false emission
line detections. To avoid this we chose to rule out APM point-like sources spectra and
only use the confirmed APM extended and paired sources. Taking into account the coverage of 
the CUYS in galactic latitude (from $\sim30^{\mathrm{o}}$ to $62^{\mathrm{o}}$), we discard 
between 1390 to 3420 K and M type stars deg$^{-2}$; these quantities were calculated
using the \cite{1980ApJS...44...73B} model. Additionally, we estimate
that nearly 4-5 QSO deg$^{-2}$ \citep{1995ApJ...441..488C} are rejected 
before any data processing effort using this method. Conversely, our method
introduces a bias against very compact ELGs: near 8 per cent of {\it star-like} ELGs were 
found in the UCM survey list 3 
sample and 11 per cent were morphologically classified as {\it compact} 
ELGs \citep{1999ApJS..122..415A}. Obviously, this bias did not affect the UCM survey. 
In absence of 
simulations better qualified than this observational fact, we
presume that 30 to 40 ELGs in the final CUYS catalogue were lost due to the selection 
effect introduced by excluding {\it star-like} ELG candidates presumably included in the 
set of point-like sources 
initially ignored. These numbers demonstrate the advantage of point-like source 
rejection as a previous step to the individual ELGs candidates processing, in spite 
of the bias described. This acquires special relevance when the follow-up spectroscopy
is not foreseen in the original observational project. 

With respect to the flux calibration, 
it was carried out using the optical continuum of 15 AGNs spectra observed
at the 2.5 m DuPont telescope by Sabbey (2001, unpublished), and included
in the CUYS catalogue, to construct the
average sensitivity curve for each column of the CCD mosaic. Such curves
were used for the flux calibration of the CUYS spectra. The sensitivity curve
reliability was double-checked using the colour-selected F8 sub-dwarf secondary
standards spectra obtained from the SDSS Data Release 2 (\citealt{2003AJ....126.2081A}, 
hereafter SDSS2) instead of AGNs spectra. These standards were originally used in the 
SDSS for their spectra calibration. Previously, we ruled out those secondary standards
with known photometric variability in periods from days to years, using data from the 
QUEST RR Lyrae Survey (\citealt{2004aj...127..1158}, hereafter QRR). The resulting
sensitivity curves are consistent (1$\sigma$) with those obtained using AGN continua. 

Finally, the extraction of raw ELG candidates followed the combination 
of three different criteria: The {\it S/N} ratio, the {\it fwhm} 
and the observed {\it EW} of emission lines on each spectrum. Several
experiments were performed to achieve optimum values and a balance 
between completeness and low contamination. The result was a step function
of observed {\it EW} and emission line {\it S/N} ratio for candidate selection.
An $EW\approx 40$\ \AA\ and emission line $S/N\approx 4$
were adopted as initial security thresholds. Above these values it is possible 
to avoid false 
emission detections \citep{1994AJ....107.1245S}.
Once cosmic
rays were discarded and additional criteria for physical emission confirmation
were applied, the spectra were subject to a final visual inspection. The completeness
analysis below demonstrates that the application of these criteria allowed us to create  
a statistically significant catalogue of ELGs.

\subsection{Survey Results}
\begin{figure}
\includegraphics[width=85mm]{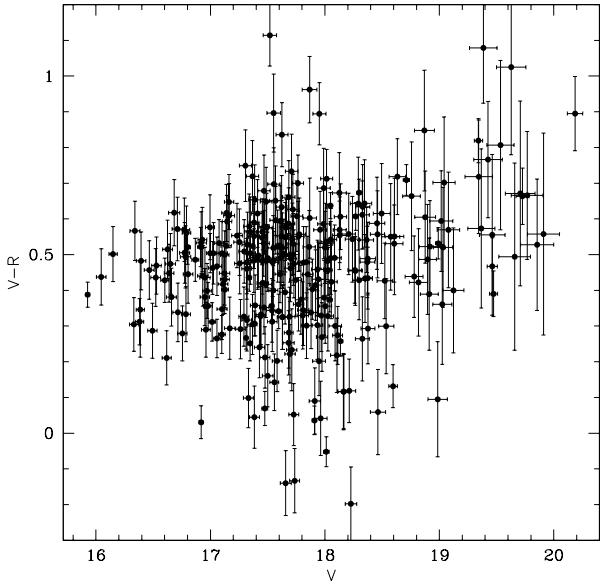}
\caption{Colour-magnitude diagram for a sample of 314 ELGs in the CUYS list.
The median brightness is $V~=~17.71$ and the median
colour $V~-~R~=~0.49$. This colour is comparable to the colour of 
local Hubble types $Im$ and $Scd$ reported by \citet{1995PASP..107..945F}.}
\label{magcol}
\end{figure}

The CUYS results are presented as a catalogue
of ELGs. The entire catalogue can be inspected in
the CIDA homepage.\footnotemark[5] 
\footnotetext[5]{Available at {\tt http://www.cida.ve/{\tiny $\sim$}abongiov/CUYS/elgcata.txt}
in ASCII format.}
It contains 427 entries and a summary of the first 10 is listed in Table~\ref{catalogue}. 
Astrometric positions 
(J2000.0) and R magnitudes of the ELG candidates were extracted from the 
USNO-A V2.0 catalogue. Instrumental photometry ($V$ and $R$ Johnson-Cousins bands) 
from QRR for 73.5 per cent (314 objects) 
of the catalogue is shown in Fig.~\ref{magcol}. The median colour $V-R$ is
0.49, a value between representative ones for Hubble types $Im$ and
$Scd$, 0.34 and 0.57, respectively \citep{1995PASP..107..945F}.

\begin{figure}
\includegraphics[width=85mm]{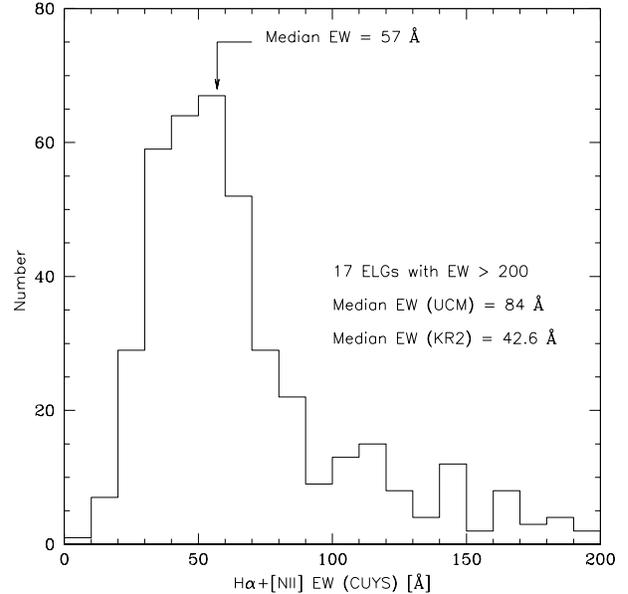}
\caption{Distribution of $\ha+[{\rm NII}]$ reduced equivalent widths for the CUYS
ELG candidates. The survey has good sensitivity for $EW\geq 30$\ \AA. The median values
for CUYS, UCM and KR2 are indicated.}
\label{histoew}
\end{figure}

\begin{figure}
\includegraphics[width=85mm]{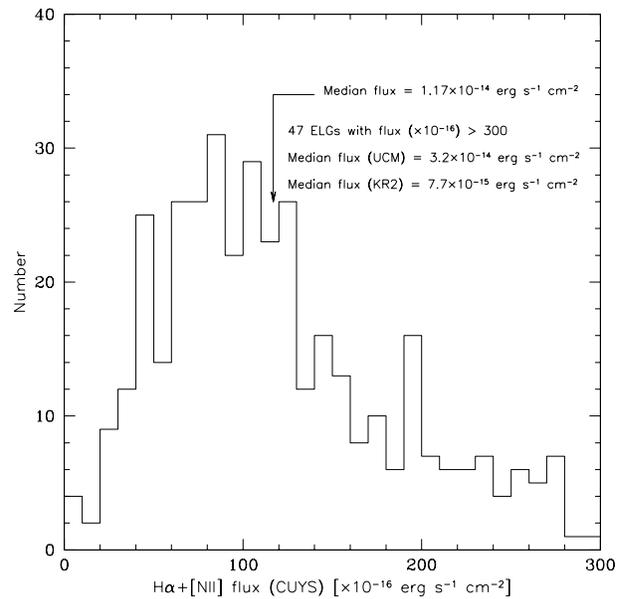}
\caption{Distribution of $\ha+[{\rm NII}]$ fluxes for the CUYS ELG candidates.
The median values for CUYS, UCM and KR2 are indicated.}
\label{histoflu}
\end{figure}

Global properties of the CUYS catalogue, in terms of {\it EW} and $\ha+[{\rm NII}]$ fluxes 
are shown in Figs.~\ref{histoew} and \ref{histoflu}, respectively. The median of
each distribution is shown together with representative values for 
the UCM \citep{1996A&AS..120..323G} and KR2 surveys. 
The median values for the CUYS catalogue are between the medians for the other surveys,
providing an objective
measurement of the {\it shallowness} of our survey relative to previous ones of
similar nature. The limitations in prism dispersion and CCD scale do not allow us to
detect ELG candidates with $EW<30$ \AA\ (see below). Thus, in this work we detect only the
contribution of the most vigorous extragalactic star-forming systems, neglecting the
bulk of ELGs with low-$EW$. Despite this handicap of our survey, we show 
that it is possible to obtain fair results with these data. 

From the original list of ELGs we selected a subset comprised by 273 objects 
belonging to the connected sky region better sampled in the survey. This
subset contains ELG candidates with emission lines blueward of 7750\ \AA\, to
avoid confusion with telluric A-band absorption in low resolution spectra.
This results in a redshift cut-off of $\rm z \sim 0.14$ for $\ha+[{\rm NII}]$ ELG. This
subset is denoted hereafter as the Uniform Subset (UnSu).  

\subsection{Comparison with SDSS results}

Taking into account that the CUYS was originally designed to study the
properties of the telescope-CCD mosaic-objective-prism combination,
a dedicated follow-up spectroscopy of ELG candidates was disregarded, in part
due to the availability of SDSS data releases after the first part of the CUYS 
was finished. We intend to make the best use of 
our observational effort and to compare the results of the CUYS survey 
with previous surveys.

\begin{figure}
\includegraphics[width=85mm]{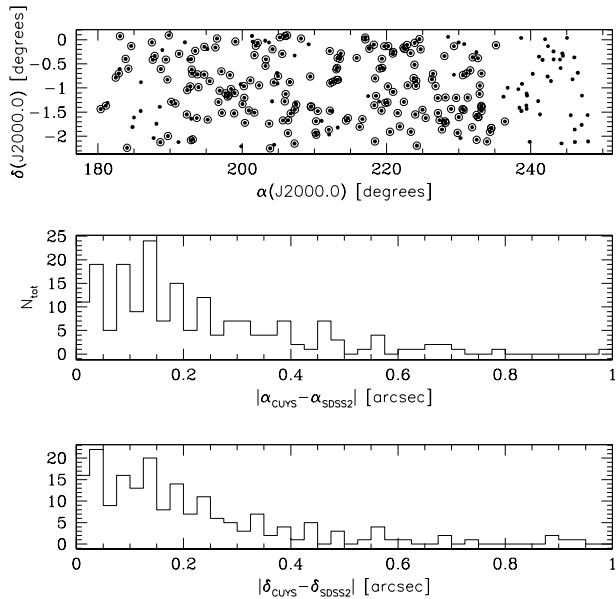}
\caption{On the top, a map with the CUYS UnSu (273 ELGs, dots).
Circled dots (190 objects) are CUYS candidates included in the SDSS2.
Objects with
$\alpha>236^{\mathrm{o}}$ (38) are not present in the spectroscopic SDSS2 dataset. 
Single dots with $\alpha\leq 236^{\mathrm{o}}$ represent CUYS UnSu objects that do not 
match SDSS2, mainly because
they are possibly too bright to be included in SDSS samples. Distributions in the other
two panels correspond to the differences in equatorial coordinates between CUYS
(USNO-A V2.0) and SDSS2 astrometry.}
\label{map1}
\end{figure}

The ELG candidates
included in the UnSu are represented in the top of Fig.~\ref{map1} (dots). About 80 per cent
of this angular extent overlaps the large sky zone explored by SDSS. From
the SDSS2 we extracted common objects (using equatorial coordinates, apparent
brightness in R band and redshift as matching parameters) in the overlapping region 
($\alpha\leq 236^{\mathrm{o}}$) as well as their spectra.
We find that all of them (190)
are extragalactic emission-line systems (including AGN), represented in Fig.~\ref{map1} by
circled dots. The histograms in this figure show the distributions of coordinates
diference between CUYS and SDSS2 objects in common: 66\% of
these differences are below $\sim$ 0.3 arcsec in both coordinates. 

There are
46 CUYS objects with $\alpha\leq 236^{\mathrm{o}}$ present in the SDSS2 imaging
database, but not in the spectroscopic database. The SDSS2 images of such objects 
were inspected individually to find the possible reasons for this exclusion. Using
the brightness ($r_{\rm pet}$) distribution of the SDSS2 objects that match with
the CUYS ELGs, we find that 16 possibly are too bright to be included in SDSS2 
spectroscopic database, 5 objects are too faint and 25 objects are complex (paired)
systems. The latter fall in the spectroscopic data hole category. Despite we do not 
have confirmation about the presence of emission lines in their spectra, these
objects are considered hereafter as $\ha+[{\rm NII}]$ ELG.  

\begin{figure}
\includegraphics[width=85mm]{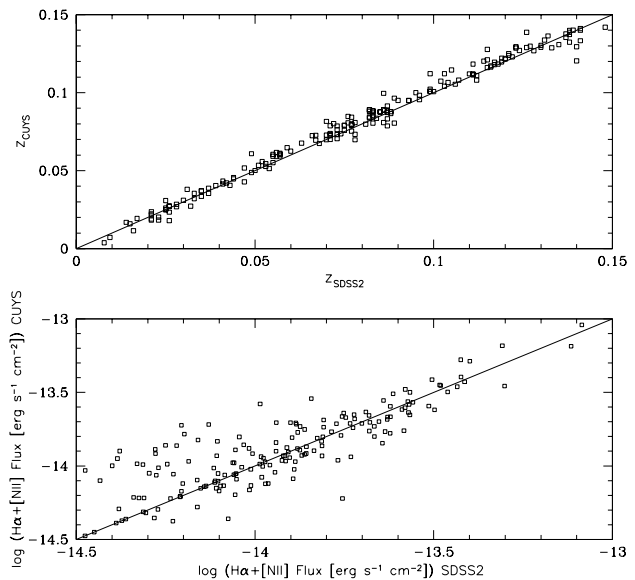}
\caption{On the top, the redshift (z$_{\rm SDSS2}$) of SDSS2 objects 
versus redshift catalogued in our survey (z$_{\rm CUYS}$). 
The rms deviation in the horizontal direction is 0.004. In the bottom panel, 
the comparison between SDSS2 and 
CUYS $\ha+[{\rm NII}]$ fluxes. The rms deviation of CUYS flux is 0.23.
In both panels, the straight line has slope$=1$.}
\label{flux_vs_flux}
\end{figure}

From the perspective of the SDSS2, 
90 per cent or more of the CUYS objects can be accepted as reliable ELGs.
A comparison of the ELG redshifts and $\ha+[{\rm NII}]$ fluxes 
between the CUYS and the SDSS2
common objects is shown in Fig.~\ref{flux_vs_flux}. The SDSS2 integrated fluxes
were obtained from the original spectra degraded to the prism characteristic dispersion
with the purpose of integrating the contribution of $\ha$ and $[{\rm NII}]6549,6584$ lines
into a single line to be fitted as the $\ha+[{\rm NII}]$ emission line of the CUYS
spectra. The correspondence is 
evidently quite good, nevertheless the intrinsic error associated with the 
CUYS absolute flux calculation is between 
4 and 8 per cent, whereas the redshift uncertainty is below 0.01. The dispersion
in the flux correlation between the SDSS2 and the CUYS common objects, especially those 
with low $\ha+[{\rm NII}]$ flux, could be due to the small and single aperture
used in the SDSS spectroscopy defined by the size of their fibers, in contrast with
the CUYS. This could explain the slight overestimation of the CUYS
flux as compared to SDSS2 $\ha+[{\rm NII}]$ flux, below $10^{-14}$ erg s$^{-1}$ cm$^{-2}$.  

On the other hand, if all $\ha+[{\rm NII}]$ emission-line systems from the 
SDSS2 (in the sky zone that overlaps the CUYS) are extracted and analysed, the results are
comparable with the CUYS in number, redshift ($z\leq0.15$), $\ha+[{\rm NII}]$ fluxes and $EW$, but 
only for $EW\geq 30$ \AA. Nearly 81\% of emission-line systems in SDSS2 (944 in this sky region) 
have $EW<30$ \AA\ (median=16.4 \AA) and the CUYS is insensitive to these objects. In other words, 
the results obtained by both surveys are nearly identical if we restrict SDSS2 to ELGs with 
$EW(\ha+[{\rm NII}])\geq 30$ \AA.  

\subsection{Survey Completeness}

The completeness of the ELG sample was determined using the $V/V_{max}$
test \citep{1968ApJ...151..393S}. A sample is considered {\it complete} below 
some apparent magnitude, 
assuming an uniform
distribution of objects, when the average $<V/V_{max}>$, defined by 

\begin{equation}
\frac{V}{V_{max}}=[\frac{r}{r_C}]^3=10^{0.6\times(m-m_C)}
\end{equation}

reaches a value of $\sim 0.5$. In this equation the subindex $C$ corresponds
to the completeness limit for which $V/V_{max}$ is being computed. 
This test indicates that the CUYS  
is statistically complete if we
include all objects in the UnSu with 
$m_{L+C}=-2.5 \log(f_{L+C})-16.32 \leq 18.05$. This subsample
of 183 objects will be called the Complete Subsample (CoSu).
The last equation defines an arbitrary magnitude scale $m\equiv m_{L+C}$ as a function
of the line + continuum flux $f_{L+C}$ (in erg s$^{-1}$ cm$^{-2}$), and the
completeness magnitude was obtained by successive aproximations of this expression. This
apparent brightness formalism was used originally by \cite{1989ApJ...347..152S}. The zero
point of $m_{L+C}$ in this survey was established using a linear fit to the 
Johnson $R$ magnitude of ELG candidates. Using this magnitude scale,
we performed an independent completeness estimate following 
\cite{1998A&AS..133..171S}, which apparently is less sensitive to space density
fluctuations in the sample. The results agree with those obtained using $<V/V_{max}>$:
the same completeness limit is reached at $m_{L+C}=18.15$, with 193 objects. 
We adopt the $<V/V_{max}>$ estimate and the results of the test are given in
Table~\ref{vvmax}. 
  
\begin{table}
\tiny
\caption{\label{t:vvmax}{$<V/V_{max}>$ test data. Column (1) contains limiting magnitude
$m_{L+C}$ adopted as described in the text. Column (2) lists the cumulative number of
ELGs brighter than $m_{L+C}$. Column (3) lists the corresponding $<V/V_{max}>$
values and column (4), the completeness percentage.}} 
\label{vvmax}
\begin{tabular}{cccc}
\hline
$m_{L+C}$&Cumulative number&$<V/V_{max}>$&Completeness\\
\hline
16.0&4&0.625&100.00 \\
16.1&4&0.544&100.00 \\
16.2&6&0.621&100.00 \\
16.3&7&0.597&100.00 \\
16.4&9&0.625&100.00 \\
16.5&9&0.544&100.00 \\
16.6&12&0.583&100.00 \\
16.7&13&0.545&100.00 \\
16.8&17&0.586&100.00 \\
16.9&24&0.636&100.00 \\
17.0&31&0.642&100.00 \\
17.1&35&0.602&100.00 \\
17.2&43&0.600&100.00 \\
17.3&54&0.609&100.00 \\
17.4&68&0.616&100.00 \\
17.5&88&0.631&100.00 \\
17.6&109&0.621&100.00 \\
17.7&126&0.594&100.00 \\
17.8&141&0.560&100.00 \\
17.9&160&0.541&100.00 \\
18.0&175&0.511&100.00 \\
18.1&190&0.483&97.94 \\
18.2&206&0.461&95.81 \\
18.3&219&0.433&93.60 \\
18.4&230&0.404&90.91 \\
18.5&239&0.375&88.85 \\
18.6&248&0.348&86.71 \\
\hline
\end{tabular}
\end{table}

Concerning the completeness of the survey at the bright end, we are missing a few
objects with $m_{L+C}<15$, whose spectra may appear saturated in our scans. This
bias does not affect our conclusions below in a statistical sense. At the faint
end, the cause of a possible deficit is discussed in the next section. 

\subsection{Catalogue Contamination}

A source of catalogue contamination is the possible presence in our sample of $z\geq0.4$
ELGs with emission features different from the $\ha+[{\rm NII}]$ blend, which are bright
enough to produce good $S/N$ ratio prism spectra. To obtain an idea about the 
fraction of these objects in the
CUYS catalogue we followed the prescription of \cite{2001ApJ...550..593J}, based
on the expected fluxes of the $\ha$, [OIII]4959,5007, H$\beta$, and [OII]3727 lines, which
could eventually be confused with $\ha+[{\rm NII}]$ in our spectra. This
analysis demonstrates, for example, that at an $\ha$ flux 
of $\sim 6\times 10^{-16}$ erg s$^{-1}$ cm$^{-2}$, the probability of finding objects 
with [OIII]4959,5007 emission in place of $\ha$ for the CUYS
spectral range is $\sim5$ per cent. Our estimate for the minimum $\ha$ flux that we detect, at least 
in the CoSu, is 
$4.62\times 10^{-15}$ erg s$^{-1}$ cm$^{-2}$, about one order of magnitude higher than the
value quoted above. Consequently, there is no evidence of contamination by lines different
to the $\ha+[{\rm NII}]$ blend in the spectra of the objects that belong to the CUYS catalogue, as 
it was confirmed using SDSS2 data.

\subsection{Spectral Classification of CUYS ELGs}

\begin{figure}
\includegraphics[width=85mm]{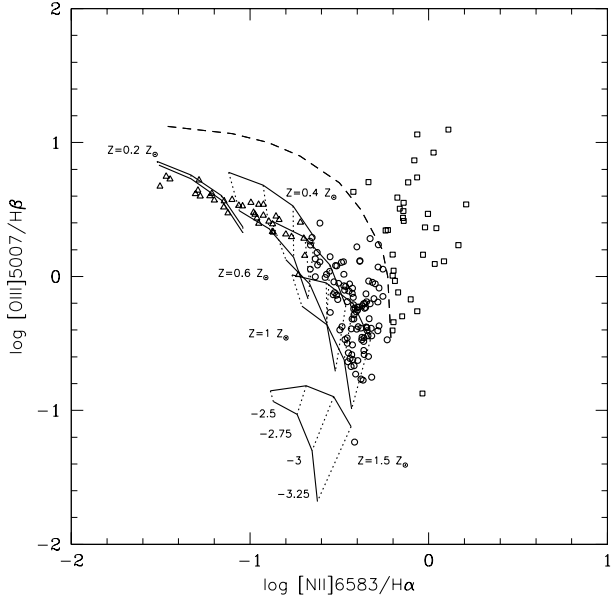}
\caption[long caption]{Diagnostic diagram $\log([{\rm OIII}]5007/{\rm H}\beta)$ versus 
$\log([{\rm NII}]6583/\ha)$ for 180 CUYS ELGs with counterparts in the SDSS2
sample. Using the prescription of \cite{1987ApJS...63..295V} and
spectroscopic criteria referred in the text, nearly
80 per cent of the CUYS sample can be regarded as HII region-like ELGs (triangles)
and SB-like ELGs (circles).
The remaining objects were classified as AGN (squares).
Solid line segments represent a sequence of models
for star-forming galaxies according to \cite{2003ApJS..149..313M}, 
as described in the text.} 
\label{veilleux}
\end{figure}

Finally, a description of the CUYS sample would not be complete without including the 
spectral classification of the
galaxies. We have separated the ELGs in three different classes: HII region-like
ELGs, SB-like (Starburst) ELGs and AGN or ``active galaxies". HII region-like
class corresponds to a synthesis of HIIH and DHIIH types
and SB-like class comprises SBN and DANS types, all described by \cite{1996A&AS..120..323G}. 
This classification was performed only for descriptive purposes and using
strictly spectroscopic criteria. Fig.~\ref{veilleux} is a representation of the diagnostic diagram
$\log ([{\rm OIII}]5007/$H$\beta)$ versus $\log ([{\rm NII}]6583/\ha)$, useful 
according to
\cite{1987ApJS...63..295V} for spectral classification of ELGs. Data for
these emission lines were extracted from 180 matched SDSS2 spectra with measurable
diagnostic lines, using the IRAF task {\tt splot}. The objects whose spectra are analyzed
here belong to the UnSu, therefore the work done here do not implies the extraction of 
{\it all} AGNs present in the CoSu, but to have a very clear idea about the {\it fraction}
of different spectral classes of ELGs present in our largest subsample. The analysis of 
the SDSS2 matched spectra 
reveals that $\sim18$ per cent of the CUYS sample is composed by AGN (squares). 
From the 143 star-forming galaxies plotted, 36 are HII region-like ELGs (triangles) 
and 107 are SB-like (circles). 

We have overplotted on the the diagnostic diagram a 
sequence of models for star-forming galaxies \citep{2003ApJS..149..313M}, to obtain
an accurate idea about metallicity and burst age of surveyed ELGs. 
The solid line segments join models of constant metallicity, whereas dotted lines 
join models of equal ionization parameter.

The sequences correspond to an instantaneous burst of star formation, with age between 0
and 2 Myr. We observe that 91 per cent of the star-forming galaxies sample lies on a locus 
with a metallicity between 0.4 and 1$\times Z_\odot$. The remaining 
objects (HII region-like ELGs)
have lower metallicities. The broad distribution of star-forming galaxies 
in this diagram can be understood as due to dispersion in the
age of the burst.

\section{CUYS \ha\ Luminosity Function in the Local Universe}
The ELG CoSu was used to derive an $\ha$ Luminosity
Function (LF) whose integral value represents an estimate of  
$\rho_{\rm SFR}$ in the Local Universe. 

To fit the LF we compute the number $\Phi$ of galaxies per unit volume
and per unit $\ha-$luminosity interval $0.4 \log L(\ha)$, given by

\begin{center}
\begin{equation}
\Phi[\log L(\ha)] = \frac{4\pi}{\Omega}\ \sum_i \frac{1}{V_i^{max}},
\end{equation}
\end{center}

where $\Omega$ is the surveyed solid angle ($\sim 0.046$ str) and $V_i^{max}$ is the volume
enclosed by a sphere of radius equal to the maximum distance the galaxy could be from
the
$ith-$object and still be detected in the survey \citep{1976ApJ...203..297S}. The sum
is performed over all the objects whose luminosity falls in the interval 
$L(\ha)\pm 0.5\ \Delta \log L(\ha)$.

Following the reasoning above, the Schechter function can be rewritten as 

\begin{center}
\begin{equation}
\Phi[\log L(\ha)]\ d\log L = \phi(L) dL,
\end{equation}
\end{center}

where $\phi(L) dL \equiv \phi^\ast (L/L^\ast)^\alpha \exp(-L/L^\ast) d(L/L^\ast)$. 
In the $V_i^{max}$ method, galaxies are assumed to be distributed homogeneously and,
for this reason, its great advantage is simplicity. Nevertheless, the data were subject to
other classical fitting schemes (\citealt{1977AJ.....82..861F}, 
\citealt{1993ApJ...404...51E}) and the results 
agree, within errors, with the ones obtained from the $V_i^{max}$ method.  

We use a median value for the ratio $(\ha+[{\rm NII}])/\ha=1.3$ derived from
the SDSS2 galaxies in common with our sample to correct objective-prism 
$H\alpha+[{\rm NII}]$ fluxes for the contribution by [NII]. This value, derived 
from original and degraded resolution
SDSS2 spectra, is consistent with the 1.33 value reported by \cite{1983ApJ...272...54K}
and \cite{1997ApJ...475..502G}, which is widely used in the literature for local ELGs.

\begin{figure}
\includegraphics[width=85mm]{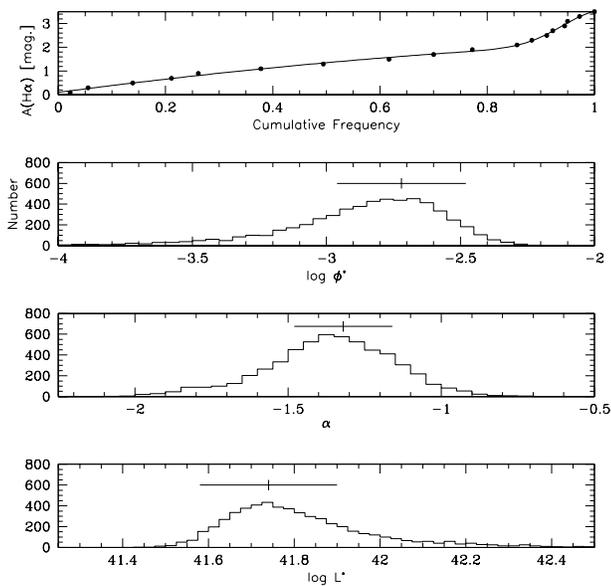}
\caption{In the top panel the distribution of reddening in $H\alpha$ from the
SDSS2 selected sample (165 ELGs) in common with the CUYS UnSu. The other plots 
contain the distribution of the Schechter function parameters obtained via Monte
Carlo experiments, with the restrictions explained in the text.}
\label{dustcor}
\end{figure}

$L(\ha)$ for each galaxy was dust corrected stochastically via $10^4$ Monte Carlo
simulations of intrinsic reddening $A(\ha)$, whose distribution is 
represented on the top of Fig.~\ref{dustcor}. The calculation of the reddening is based on 
the theoretical Balmer 
decrement $\ha/$H$\beta=2.86$ \citep{1989..O}, assuming case B recombination, $T_e=10^4$ K and 
$n_e=10^2$ cm$^{-3}$; in the case of AGNs, we used
$\ha/$H$\beta=3.10$. We take the Withford extinction curve that satisfies
$k(\lambda)=A_{\lambda}/E(B-V)$ and use the \cite{1979MNRAS...187..73P} law, which 
gives $k(\ha)=2.49$. The $A(\ha)$ distribution behaviour is a polynomial fit of
the binned inverse distribution function. 

Unlike the alternative LF(\ha) calculation
that follows in this section, we assume that the $A(\ha)$ correction probability
for each ELG in the Monte Carlo realisations is, in principle, the same. Obviously,
the $A(\ha)$ correction likelihood for each galaxy is conditioned by the shape
of the distribution function represented on the top of Fig.~\ref{dustcor}.  
The intrinsic reddening $A(\ha)$ was calculated 
using the $E(B-V)$ colour excess and {\it observed} intensity ratios $\ha_o/$H$\beta_o$
from 165 SDSS2 spectra corresponding to $\sim$61 per cent of the UnSu. In the simulations 
we also include the error distribution functions for galaxy $\ha$ flux and redshift. 

\begin{figure}
\centering
\includegraphics[width=85mm]{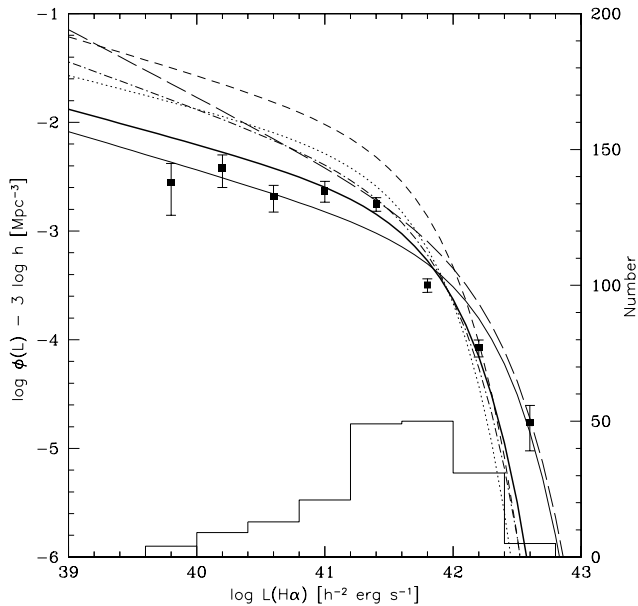}
\caption[long caption]{LF(\ha) determinations for the surveys listed in 
Table~\ref{lfcompa}. The solid thick line corresponds to the CUYS LF 
corrected for dust and errors via Monte Carlo experiments. The solid 
thin line that fits the discrete values is the CUYS LF with dust correction 
using \cite{2001ApJ...551..825J} prescription as is described in the text. 
Bars are Poissonian errors. On the bottom of the figure is the number 
distribution of CoSu ELGs used in this LF calculation. The dotted line 
corresponds to the \cite{1995ApJ...455L...1G} LF, the short-dashed line 
to \cite{1998ApJ...495..691T}, the long-dashed line to 
\cite{2000MNRAS.312..442S} and the dot-dashed line to the 
\cite{2004AJ....127.2511N} fit.}
\label{funlum}
\end{figure}

The three lower
panels of Fig.~\ref{dustcor} contain the Schechter function parameters distributions 
($\log \phi^*$, $\alpha$ and $\log L^*$, respectively) for the LF
obtained from Monte Carlo experiments (99.5 per cent confidence level). Inside each 
distribution the median and it deviation (1-$\sigma$) is represented as a bar. These 
1-$\sigma$ deviations were adopted as the corresponding error parameter in this approach, whose
representation is shown in Fig.~\ref{funlum} (solid thick line).   

\begin{figure}
\includegraphics[width=85mm]{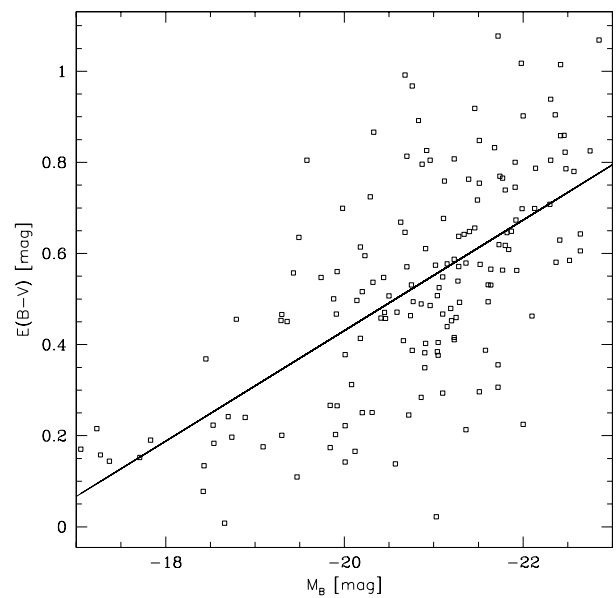}
\caption{Relation between colour excess $E(B-V)$ and absolute magnitude $M_B$ for
the 165 CUYS objects that match SDSS2 galaxies. 
The linear
fit (residual 0.19 mag rms) was used to calculate $A(\ha)$ for each galaxy of the UnSu before 
performing the LF(\ha) alternative estimation.} 
\label{corrmbebv}
\end{figure}

We also implemented a simpler form to correct the $L(\ha)$ by reddening. It 
consists in deriving a correlation of $E(B-V)$ with absolute magnitude $M_B$ just as presented
by \cite{2001ApJ...551..825J}, but using the data from the 165 SDSS2 galaxy spectra.
A linear fit 
between these variables is shown in Fig.~\ref{corrmbebv}. The rms of the residuals reaches
$\sim$0.2 mag in color excess, which represents a \ha\ extinction of about 0.6 mag. 
We agree with these authors in the scatter in this trend. In spite of this, the
correlation was used to correct $L(\ha)$ for each galaxy individually and the
parameters of the LF(\ha) were calculated using a Marquardt-Levenberg
least-squares fit: $\log L^{*}=42.06\pm0.14$, $\alpha=-1.35\pm0.16$ and 
$\log\phi^{*}=-3.12\pm0.25$. This LF(\ha) is represented in Fig.~\ref{funlum} by a 
solid thin line. The number distribution of CoSu objects used in this LF
calculation is shown in the bottom of the figure. The error bars in the
points are Poissonian errors.

Despite the relative shallowness of the CUYS, the parameter values and associated
errors obtained with both approaches are acceptable. The stochastic method favours
slightly the fainter luminosities whereas the alternative approach, the brighter ones.
We adopt the parameters resulting from stochastic aproximation to the LF(\ha) calculation
problem because it takes into account the error distribution functions for
galaxy $\ha$ flux and redshift. The resulting LF parameters from this method are 
listed in Table~\ref{lfcompa} together with values
for different surveys performed after the UCM survey \citep{1995ApJ...455L...1G} for 
comparison purposes, and all of them shown in Fig.~\ref{funlum}.
A number defect in CUYS LF($\ha$) at fainter
luminosities is evident when it is compared with previous estimations. This is 
evidently related to the CUYS selection effects, mainly the absence of ELGs with 
$EW<30$\AA\ and, to a smaller extent, the exclusion of point-like ELGs  
as part of the data reduction process.  

\begin{table*}
\centering
\begin{minipage}{180mm}
\caption{\label{t:lfcompa}CUYS LF(H$\alpha$) and related data compared with earlier works 
as tabulated in \citealt{2004AJ....127.2511N} ($h=1$). The $\log \cal L$(H$\alpha$) value
for the CUYS {\it includes} the contribution by AGNs.} 
\label{lfcompa}
\begin{tabular}{lrrrrr}
\hline
&\cite{1995ApJ...455L...1G}&\cite{1998ApJ...495..691T}&\cite{2000MNRAS.312..442S}&\cite{2004AJ....127.2511N}&{\bf CUYS}\\
\hline
Survey area & 471.4 deg$^2$ & 500 arcmin$^2$ & $\sim10\ $ deg$^2$   & 229.7 deg$^2$ & 151.1 deg$^2$\\
Mean redshift & $\sim 0.025$         & $0.21$          & $0.15$             & $0.054$       & $0.081$\\
Size of the sample & 176            & 110               & 159                & 665          & 183\\
$\log L^{*}$ & $41.56\pm0.08$          & $41.61\pm0.13$    & $42.11\pm0.14$     & $41.68\pm0.10$  & $41.74\pm0.16$\\
$\alpha$     & $-1.3\pm0.2$             & $-1.35\pm0.06$    & $-1.62\pm0.10$     & $-1.43\pm0.10$ & $-1.32\pm0.16$\\
$\log\phi^{*}$ & $-2.3\pm0.2$             & $-2.09\pm0.09$    & $-3.04\pm0.20$     & $-2.56\pm0.30$ & $-2.72\pm0.24$\\
$\log \cal L$(H$\alpha$) & $39.38\pm0.04$ & $39.66\pm0.04$    & $39.43\pm0.06$     & $39.31\pm0.04$ & $39.23\pm0.07$\\
\hline
\end{tabular}
\end{minipage}
\end{table*}

Given a Schechter function aproximation to the \ha\ 
luminosity distribution in the volume of local universe considered, the integrated 
$\ha-$luminosity $\cal L$(H$\alpha$) has the value

\begin{center}
\begin{equation}
{\cal L}({\rm H}\alpha) = \int_0^\infty \phi(L)\ L\ dL = \phi^*L^*\Gamma(2+\alpha),
\end{equation}
\end{center}

which is given in the last line of the Table~\ref{lfcompa}. All the estimates in Table~\ref{lfcompa}
seem very similar, the CUYS $\cal L$(H$\alpha$) is the smallest value, but it is very close
to the estimate of \cite{1995ApJ...455L...1G}. Probably, our number defect of ELGs 
is real in the volume of universe surveyed. We need to stress here 
again that the CUYS value for $\cal L$(H$\alpha$) is valid for star-forming 
galaxies with $EW(\ha+[{\rm NII}])>30$\AA. 

We translate the integrated $\ha-$luminosity $\cal L$(H$\alpha$) into a SFR density via
the transformation

\begin{center}
\begin{equation}
{\cal L}({\rm H}\alpha) = 1.21\times 10^{41}\ \rho_{\rm SFR},
\end{equation}
\end{center}

which is taken from \cite{2003ApJS..149..313M}, assuming a Salpeter IMF from 0.1 to
125 M$_\odot$, and solar stellar and gas metallicity. Consequently, 
the estimate from the CUYS is (statistical error)

\begin{center}
\begin{equation}
\rho_{\rm SFR} = 0.014\pm0.002\ h\ {\rm M}_\odot\ {\rm yr}^{-1}\ {\rm Mpc}^{-3}.
\end{equation}
\end{center}

Obviously, this quantity must
be regarded with caution. In the CUYS LF calculation we
{\it included} AGNs. As it was explained in the previous section, we are sure about 
the {\it fraction} of these objects in the sample, but not
on their identification all over the catalogue. We preferred to leave them in the 
sample rather than extract 
only a fraction. Additionally, we do not correct the \ha\ flux for stellar absorption.
This correction is really negligible given the 
$EW$ values considered here. Thus, if we use the table with LF(H$\alpha$) systematic 
errors calculated 
by \cite{2004AJ....127.2511N},
the CUYS $\rho_{\rm SFR}$ value has a 
combined systematic error 
of at most $\pm0.003\ h\ $M$_\odot\ $yr$^{-1}\ $Mpc$^{-3}$ due to neglecting stellar 
absorption and the AGNs inclusion.

\section{Spatial Distribution of CUYS ELG}

The CUYS provides a good opportunity to obtain information
about the spatial 
distribution of ELGs with high SFR via 2-point correlation functions
and to establish a preliminary determination of the 
ELG sample bias compared to a recent set of ``normal" or ``quiescent" galaxies
in the same spatial volume. Below we study
the relation between star formation and environment
conditions from three different perspectives.

\subsection{Reference Catalogue and Qualitative Environmental Effects}

\begin{figure}
\includegraphics[width=90mm,angle=90]{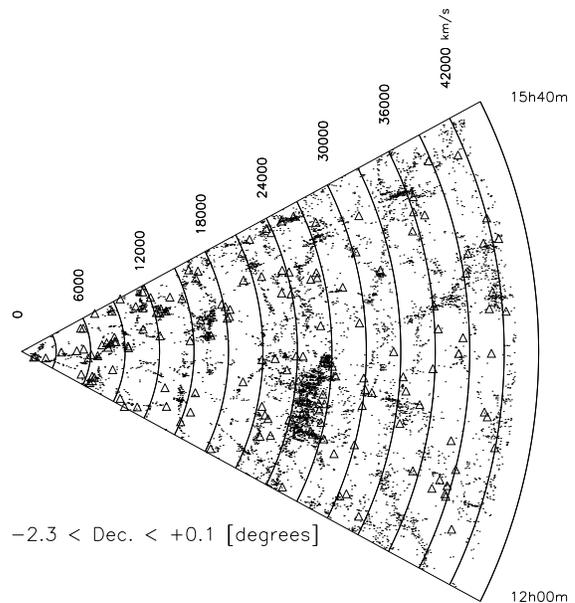}
\caption{Pie diagram with RC1 galaxies (small dots; 9,364 objects) and CUYS CoSu 
ELG (triangles; 161 objects) that lie in the same volume.}  
\label{pie}
\end{figure}

First, we wish to compare the CUYS ELG spatial distribution with a deeper sample
of galaxies than the usual bright galaxy catalogues. As a first reference catalogue (RC1)
we use a subset of 9,364 quiescent galaxies from the spectroscopic sample of SDSS2, 
with $R\leq18.15$ and $z\leq0.14$ as the selection criteria. The completeness limit of 
the SDSS galaxy catalogue
to $z\sim0.1$ is $r_{\rm pet}\leq17.8$ \citep{2003ApJ....592..819}. This
selection resembles the CUYS sample in terms of limiting
flux and volume of universe. From the RC1 we extracted galaxies that match the CUYS CoSu.
Unfortunately, the
angular overlap between both catalogues forced us to limit the CUYS CoSu to 
132 deg$^2$ (161 objects). A representation of the RC1 and the CUYS CoSu 
(trimmed) in velocity space can be seen in Fig.~\ref{pie}. It is easy to distinguish
density enhancements in the distribution of RC1 galaxies in redshift space. To obtain an
estimate of the RC1 galaxy clustering, we use the \cite{1982ApJ...257..423H} 
{\it Friends-of-Friends} ({\it FoF}) algorithm for
group and cluster finding in magnitude-limited samples of galaxies 
(with $D_0= h^{-1}\ 0.4$ Mpc, the projected separation chosen
at some fiducial redshift and $V_0=350$ km s$^{-1}$, adopted as parameters
after various trials). We find 307 groups with between 5 and 350 members each. We discarded
groups with less than 5 members. Once the group
and cluster catalogue was constructed, we divided the CUYS CoSu in two subsets: {\it grouped},
associated with one or more quiescent galaxy groups (17 objects), and {\it isolated} ELGs 
(144 objects). We find a ratio $SFR_{isolated}/SFR_{grouped}=1.4\pm0.3$.
If we use the CfARS catalogue from Huchra \etal\ (2001, private communication)
for double-checking this result, this ratio reaches $2.1\pm0.5$. 
The CfARS catalogue is complete
to $B=15.24$ and the sample used for our purposes has 669 galaxies. It seems clear 
that the average star 
formation in low density environments is
about a factor of 1.5 to 2 greater than in group and cluster regions, at least when
galaxies with copious star formation are used as tracers of the SFR. 

\subsection{2-point Correlation Functions}

We used the RC1 and the CUYS CoSu (trimmed) sets of galaxies to examine clustering
properties by means of auto-correlation algorithms in redshift space. With this purpose,
we adopt the correlation function (CF) formalism proposed by \cite{1993ApJ...417...19H}. 
This correlation function estimator is defined by

\begin{center}
\begin{equation}
\xi(s) = {\langle NN(s) \rangle \langle RR(s) \rangle \over {\langle NR(s) \rangle}^2} -1,
\end{equation}
\end{center}

where $NN(s)$ is the number of catalogue galaxies in the interval $[s,s+ds]$, 
$RR(s)$ is the number of pairs in the same interval in the random 
catalogue ($\sim20$ times larger than the size of data sample), and $NR(s)$ is 
the number of pairs in the combined sample
with this separation.
The angular brackets in this equation denotes
average over all pairs separated by $s$ in the galaxy sample.
  
The error in the CF estimate is 
defined as the standard deviation at each point where the CF 
was measured. To diminish boundary effects, the RC1 sample was enhanced by 2 degrees in
angular dimension and the random catalogues were re-sampled in each simulation.

It was possible to obtain positive values of the 2-point CF for the
CUYS CoSu only when
$s > h^{-1}\ 1$ Mpc. This indicates that the ELGs are anti-clustered, or
that they are arranged regularly in redshift space at smaller scales. 
The galaxy correlation function can be approximated by a
power-law \citep{1980..P}

\begin{figure}
\includegraphics[width=85mm]{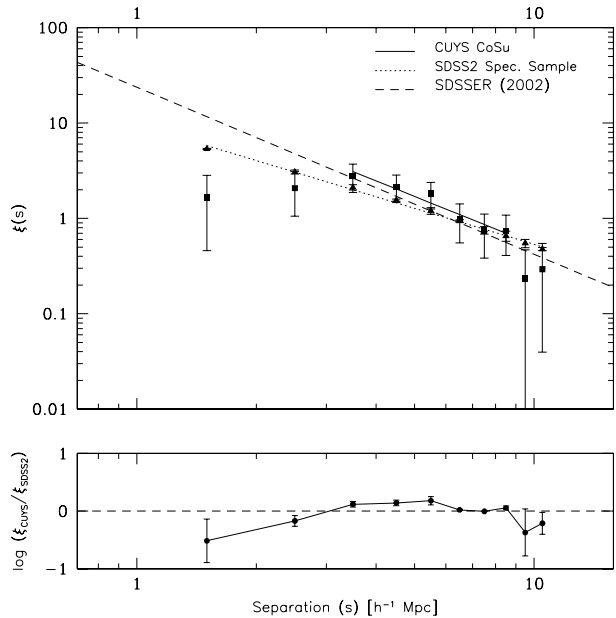}
\caption{The top panel shows the redshift space 2-point correlation function for 
the CUYS CoSu
(squares) and the SDSS2 sample (triangles). The bars show the standard deviation. The
continuous line represents the CUYS CF in the scale range $3.5-8.5\ h^{-1}\ $Mpc.
The dotted line is the fit to CF for the SDSS2 sample and the dashed line is a reference
from the SDSSER (2002). The bottom panel shows the bias between the CUYS CoSu
and the SDSS2 sample correlation function, defined as the quotient of both quantities.}
\label{cf}
\end{figure}

\begin{center}
\begin{equation}
\xi(s) = \left(\frac{s}{s_0}\right)^{-\gamma},
\end{equation}
\end{center}

where $s_0$ is the correlation length (defined as $\xi(s_0)\equiv 1$) and $\gamma$
is the power-law slope. The top panel of Fig.~\ref{cf} shows
estimates of the redshift space 2-point CF
for each sample. The results of the fits are given in Table~\ref{cfs}.

The points in Fig.~\ref{cf} are distributed in 1 $h^{-1}\ $Mpc bins 
reaching up to $\sim 11\ h^{-1}\ $Mpc, where the known
shoulder in $\xi(s)$ is not yet appreciable. Moreover, beyond this limit, the CF 
determination for the CUYS CoSu was unacceptably noisy. 
We include for reference in Table~\ref{cfs} and Fig.~\ref{cf} the fit obtained
by \cite{2002ApJ...571..172Z} for general galaxy clustering in the SDSS Early Release (SDSSER).
A power-law fitting for the CUYS CoSu was possible only 
in the scale range 3.5-8.5 $h^{-1}\ $Mpc.

It is evident that the clustering amplitude 
in this scale range is similar in the 3 samples. ELGs appear to be practically as clustered as 
SDSS2 galaxies. On the other hand, below $3.5\ h^{-1}\ $Mpc, ELGs seem 2 to 5 times
less clustered than quiescent galaxies. The bias pictured 
in the bottom panel of Fig.~\ref{cf}, defined as the ratio $\xi_{\rm CUYS}/\xi_{\rm SDSS2}$,
shows an asymptotical behaviour on scales $s>3\ h^{-1}\ $Mpc with an apparent shoulder on smaller
scales that should be confirmed with larger ELG samples.

\begin{table}
\caption{\label{t:cfs} Correlation function parameters ($h=1$).}
\label{cfs}
\begin{tabular}{lccc}
\hline
Sample&$s_0$ [Mpc]&$\gamma$&scale range [Mpc]\\ \hline
CUYS CoSu&6.87$\pm$0.79&1.68$\pm$0.22&3.5-8.5 \\
SDSS2&6.08$\pm$0.05&1.26$\pm$0.06&1.5-10.5 \\
SDSSER (2002)&6.10$\pm$0.03&1.75$\pm$0.03&0.1-16.0 \\
\hline
\end{tabular}
\end{table}

\subsection{Environment and SFR density}

As indicated in the introduction to this paper, a recent and detailed analysis
about the effect of environment on the star formation activity in the local universe
was published by \cite{2003ApJ...584..210G}. They found that the SFR in nearby
galaxies is strongly correlated with the {\it projected} galaxy density in at least three diferent
(and complementary) ways: (a) the decrease of the overall SFR distribution in dense environments
compared to the field population; (b) this effect is most noticeable for the stronger star-forming
galaxies; (c) the existence of a characteristic density in the density-SFR
relation at a local galaxy density of $\sim 1\ h_{75}^{2}$ Mpc$^{-2}$.  

Despite the limitations imposed by the absence of a morphological classification of CUYS 
CoSu ELGs and the relative small size of this subsample, we have extracted
additional information
about the effects of environment on SFR density. 
The availability of a characteristic correlation lenght $s_0$ for the CUYS ELGs give
us the opportunity to use a different approach in the quantification of the 
environmental effects discussed in the section 4.1. The distance $s_0$
separates the regime of large spatial density fluctuations
from the regime of small fluctuations. The latter can be interpreted as a 
homogeneous galaxy distribution, but this is a current topic of debate 
\citep{1999ApJ...522L...5G}. 

\begin{figure}
\includegraphics[width=85mm]{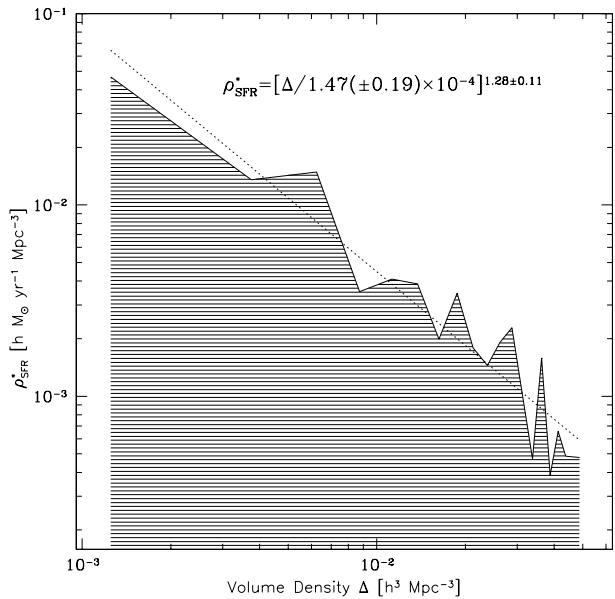}
\caption{Relation between the volume density of galaxies and the star formation rate
density inside spheres of radius equal to the correlation length $s_0$ in
redshift space derived from the 2-point correlation function fitted 
to the CUYS CoSu sample.}
\label{sfrclust}
\end{figure}

We now use the CUYS CoSu correlation length $s_0$ as 
a characteristic radius to compute the environment galaxy density in the
redshift space. This corresponds to an {\it ELG-centric} perspective
of the environmental effects on star formation. We calculate the average 
density of galaxies, $\Delta$, and the
star formation rate density $\rho^*_{\rm SFR}$ inside spheres of radius $s_0$
centered in each galaxy of the CUYS CoSu, integrating the $L(\ha)$ of the galaxies 
involved in each volume. The results are expressed as a
distribution function in Fig.~\ref{sfrclust}. The 
inverse proportionality between the environment and the SFR 
spatial density is not only evident, but a power-law behaviour describes
rather well this phenomenon. The best-fit to the distribution is 

\begin{center}
\begin{equation}
\rho^*_{\rm SFR}\ (h=1)=\left(\frac{\Delta}{1.47(\pm0.19)\times10^{-4}}\right)^{1.28\pm0.11}. 
\end{equation}
\end{center}
  
This is the first reported quantitative parametrisation
of the SFR density dependence on the average {\it spatial} density of galaxies based on 
ELGs with vigorous star formation. We have deliberately omitted the separation
between dense environments and field populations of ELGs to obtain
a perspective about the {\it continuum} without ``breaks'' that seems to link both
variables at a distance scale comparable to $s_0$. An interesting future work
could be to profit from the SDSS recent releases to explore 
more deeply the spatial environmental effects on the SFR, using perhaps other 
descriptors of galaxy clustering.  

\section{Conclusions}
In this paper we present the basic properties of the objective-prism CUYS 
for ELGs, with special
emphasis on the z$\leq0.14$ sample . The size, covered sky area, and 
$\ha+[{\rm NII}]$ flux distribution, make our survey competitive with
other modern ELG surveys. Technical limitations forced us to search and
find ELGs with $EW(\ha)>30$ \AA. Thus, all the work in this paper refers to
local galaxies with vigorous star formation. Our main results can be summarized
as follows:

\begin{itemize}
\item Objective-prism searchs for ELGs provide a productive and cheap method to
generate large catalogues of local SB-like, HII region-like and active galaxies. 
We used the driftscan technique, coadding 1-dimensional reduced spectra from
multiple scans of the same region of the sky.
\item The typical colour of ELGs in our sample is $V-R=0.49$, comparable to the
mean colour of local Hubble types {\it Im} and {\it Scd}. The analysis of
the diagnostic diagram for spectral classification reveals that more than
80 per cent of the CUYS ELGs can be considered star-forming galaxies. From
this fraction, nearly 75 per cent is dominated by SB-like ELGs. Photoionization
models imply that these galaxies have metallicities
between 0.4 and 1$\times Z_\odot$ with instantaneous bursts of star formation
aged less than 2 Myr.
\item Compared with the LF($\ha$) derived from deeper surveys, the CUYS LF($\ha$) shows good
agreement with previous results except in the low $\ha$ luminosity region,
where we find a small deficit attributed to the survey selection effects. 
Nevertheless the star formation rate density that we calculate
is in agreement with values reported in the literature. 
\item On average, the SFR for galaxies associated with
groups/clusters of quiescent galaxies is 0.5 to 0.7 times the value
for isolated ELGs. The fraction of ELGs regarded as {\it grouped} versus 
{\it isolated} is 0.1; additionally, the mean density ratio between star-forming
and quiescent galaxies is on the order of $\sim2.5$. Thus, it is clear that ELGs
avoid dense regions of quiescent galaxies.
\item ELG clustering is statistically indistinguishable from quiescent galaxy 
grouping in redshift space scales from 3.5 to 8.5 $h^{-1}\ $Mpc. 
Between $\sim1$ and 3 $h^{-1}\ $Mpc, 
ELGs are 2 to 5 times less clustered than quiescent galaxies, whereas
below 1 $h^{-1}\ $Mpc ($\sim$1 Abell radius) in redshift space, ELGs are 
anti-clustered or they are arranged regularly. This result is in agreement with
the hierarchical galaxy formation scenario. Perhaps today ELGs are isolated
systems that have not yet started the merging process. Conversely, merged galaxies
that formed isolated luminous galaxies or today cluster galaxies, have been depleted
of enormous amounts of gas and they show today few star formation
events.
\item Finally, we propose a parametrization of the relation 
between SFR density and environment density. As far as this survey is concerned, there 
seems to exist a continuum dependence between these variables which can be described by
a power-law.   
\end{itemize}

\section*{Acknowledgments}

We are especially grateful to CIDA colleagues who have participated in different
discussions about topics directly involved in this paper. Special thanks
to the night assistants O.~Contreras, F.~Moreno, R.~Rojas and U.~S\'anchez and 
the professional observers L.~Romero and D.~Herrera of the Venezuelan
National Astronomical Observatory for their constant effort.

We also acknowledge the useful suggestions from an anonymous referee. The alternative
calculation of the LF(\ha) using an extintion correction estimated from a correlation 
between $M_B$ and $E(B-V)$ is included in the referee's recommendations.

The observations reported in this paper were done under the  
Scientific and Academic
Collaboration Agreement signed between the
Universidad Complutense de Madrid (UCM), Espa\~na, and the Centro de Investigaciones
de Astronom\'{\i}a ``Francisco~J.~Duarte" (CIDA), Venezuela, in force since 
1996 until end of 1999.

Funding for the Sloan Digital Sky Survey (SDSS) has been provided
by the Alfred P. Sloan Foundation, the Participating Institutions, the National
Aeronautics and Space Administration, the National Science Foundation, the U.S.
Department of Energy, the Japanese Monbukagakusho, and the Max Planck Society.
The SDSS Web site is {\tt http://www.sdss.org/}. The SDSS is managed by the Astrophysical
Research Consortium (ARC) for the Participating Institutions. The Participating
Institutions are The University of Chicago, Fermilab, the Institute for Advanced
Study, the Japan Participation Group, The Johns Hopkins University, Los Alamos
National Laboratory, the Max-Planck-Institute for Astronomy (MPIA), the
Max-Planck-Institute for Astrophysics (MPA), New Mexico State University,
University of Pittsburgh, Princeton University, the United States Naval
Observatory, and the University of Washington.
 
A.B. acknowledges R.~Guzm\'an (University of Florida, USA), A.~Parravano (Universidad
de Los Andes, Venezuela), M.~Bautista (Instituto Venezolano de Investigaciones
Cient\'{\i}ficas), M.~Bessega (Universidad Central de Venezuela), J.~Bergamaschi
(Observatorio Cagigal, Caracas),
who provided useful advice on the analysis of survey results.

A.B. also acknowledges financial support from the Fondo Nacional
de Ciencia, Tecnolog\'{\i}a e Innovaci\'on, Ministerio de
Ciencia y Tecnolog\'{\i}a, Venezuela.

\label{lastpage}

\end{document}